\documentclass[12pt]{article}
\usepackage{graphicx}

\makeatletter
\newcounter{subeqncnt}
\def\thesubeqncnt{\alph{subeqncnt}}
\def\subequations{\begingroup
\stepcounter{equation}\edef\@tempa{\theequation}
\let\c@equation\c@subeqncnt\c@subeqncnt\z@
\edef\theequation{\@tempa\noexpand\thesubeqncnt}}

\makeatother

\begin{document}
\baselineskip=5.0ex

\begin{center}
{\large\bf Optimal damping algorithm for unrestricted Hartree-Fock calculations}
\par
\medskip
{Jun-ichi Yamamoto$^a$, 
Yuji Mochizuki$^{b,c}$\footnote{Corresponding author: Fax +81 3 3985-2407, fullmoon@rikkyo.ac.jp}}
\par \smallskip
$^a${\it HPC Division, NEC Corporation, 1-10 Nisshin-cho, Fuchu-shi,
Tokyo 183-8501, Japan} \par
$^b${\it Department of Chemistry and Research Center for Smart Molecules, 
Faculty of Science, Rikkyo University,
3-34-1 Nishi-ikebukuro, Toshima-ku, Tokyo 171-8501, Japan} \par
$^c${\it Institute of Industrial Science, The University of Tokyo, 4-6-1 Komaba, 
Meguro-ku, Tokyo 153-8505, Japan} \par

\medskip
{\bf 2013/2/25 - JST} \par

\end{center}

\newpage
\noindent{\Large\bf Abstract} \par
We have developed a couple of optimal damping algorithms (ODAs)
for unrestricted Hartree-Fock (UHF) calculations of open-shell molecular systems. 
A series of equations were derived for both concurrent and alternate
constructions of $\alpha$ and $\beta$ Fock matrices
in the integral-direct self-consistent-field (SCF) procedure.
Several test calculations were performed to check the convergence behaviors. 
It was shown that the concurrent algorithm provides better performance
than does the alternate one. \par

\bigskip
\noindent{\Large\bf Keywords} \par
Open-shell; Unrestricted Hartree-Fock (UHF); Optimal Damping Algorithm (ODA);
Self-Consistent-Field (SCF)
\par

\newpage
\bigskip
\noindent{\Large\bf 1. Introduction} \par
The Hartree-Fock (HF) method has long been the fundament in molecular orbital 
(MO) calculations. In 1951, Roothaan [1] developed systematic operator 
formulations known as the restricted HF (RHF) method for the ground state 
of closed-shell systems where each occupied orbital
has a pair of $\alpha$- and $\beta$-spin electrons. 
To treat radicals with open-shells, Pople and Nesbet [2] provided 
the unrestricted extension (UHF) by which $\alpha$-spin orbitals 
and $\beta$-spin orbitals can be different by incorporating 
the spin polarization at the cost of spin contamination. 
Roothaan [3] successively proposed the restricted version 
for open-shell molecules (ROHF) without the contamination problem, 
and afterward Plakhutin et al. [4] remade the operator formalism 
of ROHF in a canonical style. Recently, the close relationship 
between UHF and ROHF was re-evaluated by Tuchimochi and Scuseria [5,6]. \par

The HF equations should be solved under the self-consistent-field (SCF)
condition with nonlinear dependence through the density matrix. 
However, it is well recognized that the simple iterations suffer from 
the convergence difficulty even for closed-shell cases [7-9]. To reduce the
difficulty, there were two older damping techniques proposed by Saunders and 
Hiller [10] (level shift) and by Zerner and Hehenberger [11] 
(dynamic damping). Then, Pulay [12,13] invented a breakthrough 
approach to improve the convergence of HF SCF procedure, 
named as the direct inversion in the iterative subspace (DIIS).
DIIS was designed to minimize the squared norm of residuum under a 
normalization constraint. After Pulay's success, various related variants 
were developed [14-21], where these DIIS methods were usually formulated
for the Fock and density matrices with atomic orbital (AO) basis
functions to expand MOs. Particularly, C2-DIIS by Sellers [15]
as well as Energy-DIIS (EDIIS) by Kudin et al. [16] have been used 
most widely. The latter has a connection to the optimal damping
algorithm (ODA) derived by Canc\`es and Bris [22,23], 
a direct minimization technique of RHF energy
under the relaxed constraints of idempotency. \par

In the direction of MO-oriented SCF optimization,
Bacskay [24] pioneered the quadratically convergent SCF (QC-SCF) procedure
through the second-order Newton optimization of the RHF energy. 
In QC-SCF, the occupied MOs should be improved by the explicit mixing
with the unoccupied MOs without any Fock matrix diagonalization
with respect to AO-indices during the iteration. However, the actual 
computations could be too costly
due to the explicit evaluation of orbital Hessian. Anyhow, the diagonalization-free 
nature of MO optimization is favorable to the integral-direct 
SCF calculation [25] with parallelism [26]. Hence, several less expensive
procedures with efficient approximations [27-31] have thus been devised. 
An AO-based Newton technique [32] was developed as well. \par

As pointed out in Ref. [7], UHF could easily encounter 
the convergence problem. Claxton and Smith [33] reported the direct minimization
recipe for improvement. Seeger and Pople [34] proposed an MO-based optimization 
approach for UHF, while Bacskay [35] extended his QC-SCF [24] for the ROHF case.
Neese [36] revised the approximated second-order SCF (SO-SCF) method
of Ref. [30] by adding the $\alpha$-$\beta$ coupling elements. In comparison with
the case of closed-shells or RHF, works oriented to the calculations of 
open-shell (UHF or ROHF) have been rather limited. \par

In this paper, we propose a couple of ODAs designed for UHF calculations 
[2,9]; UHF could still be a reasonable zeroth-order treatment for open-shell
systems as long as the spin contamination is small enough.
Canc\`es and Bris [23] certainly addressed the application of their 
ODA/RHF to the UHF case. Nonetheless, both corresponding formulation 
and numerical result were not shown. We present here
detailed formulations and algorithmic descriptions of ODA/UHF, 
which should be useful for further methodological developments.
Our proposal covers two ways of AO-based Fock matrix
construction in the integral-direct SCF procedure [25], by
incorporating several $\alpha$-$\beta$ coupling terms.
One is the alternate construction in which Fock matrix
for each spin (say $\alpha$) is computed at a certain 
step of iteration and then the corresponding density matrix is updated
for the construction of another spin ($\beta$).
The other is the concurrent construction where both $\alpha$
and $\beta$ Fock matrices are simultaneously computed
and $\alpha$ and $\beta$ density matrices are then 
updated in a single step; the cost of direct integral
generations is a half of that for the alternate construction.
An attractive point of ODA is 
a relatively small requirement of memory resource. The remaining part 
of this paper is organized as follows. In Section 2, a brief summary of the ODA/RHF 
method in Ref. [23] is given for self-completeness and
later convenience. Section 3 describes two ODAs for UHF in detail. 
In Section 4, test applications with four examples, e.g. CN radical, are shown. \par

\bigskip
\noindent{\Large\bf 2. Brief summary of ODA/RHF} \par
Since the basic formulation and notation of our ODA/UHF follow 
the original ones of ODA/RHF by Canc\`es and Bris [23], 
the essential equations are summarized in this SubSection. 
For simplicity, all matrices are written in Capital 
italic font (or without bold font) hereafter.
The subscript specifies the step number of SCF iteration. 
The total number of electrons is $2N_e$ in the closed-shell RHF
description. The number of AOs (or dimension of matrix) is $N$ which
specifies the formal dimension of matrices. \par

The Fock matrix $F$ is given as [9]
\begin{eqnarray}
F = h + G(D), \\ % (1)
G(D) = 2J(D) - K(D), \\ % (2)
D = C C^*. % (3)
\end{eqnarray}
Here, $h$ is the one-electron contribution (kinetic energy and nuclear
attraction energy), and the two-electron contribution $G$ (consisting
of Coulomb $J$ and exchange $K$) has the dependence on the density 
matrix $D$ formed from the AO-MO coefficient matrix $C$; asterisk 
corresponds to transposition. 
$C$ is obtained by solving the general eigenequation
\begin{eqnarray}
F C =SC \varepsilon , % (4)
\end{eqnarray}
where $S$ is the overlap matrix and $\varepsilon$ 
contains the orbital energies in the diagonal elements. 
$D$ of Eq. (3) satisfies two crucial constraints associated 
with the orthonormal condition
\begin{eqnarray}
N_e={\rm Tr} \left[ DS \right] ~~~~{\rm (Number~ of~ electron)}, \\ % (5)
DSD=D ~~~~{\rm (Idempotency)}. % (6)
\end{eqnarray}
The RHF electronic energy is then written as
\begin{eqnarray}
E^{\rm RHF} (D) = {\rm Tr}\left[ 2 h D +  G(D) D \right]  =
{\rm Tr} \left[ h D  +  F(D) D \right]  . % (7)
\end{eqnarray}
$C$ is updated through the iterative SCF optimization until 
the convergence criteria involving energy and density
are satisfied under the given thresholds. \par

The $k$-th step of SCF iteration consists of
\begin{enumerate}
 \item Assemble Fock matrix $F_k=h+G(D_k)$,
 \item Compute RHF energy $E_{k}= {\rm Tr}\left[ h D_k +  F_k D_k \right]$,
 \item Solve eigenvalue problem $F_kC_k= SC_k \varepsilon_k $,
 \item Form density matrix $D_{k+1}=C_kC_k^*$,
 \item Check convergence; go to $k+1$-th step if not converged.
\end{enumerate}
As already denoted in Section 1, the above listed procedures are generally 
slow to converge [7-9], and thus a variety of 
acceleration techniques are highly necessary. \par

Now, the ODA/RHF procedure is briefed according to Ref. [23].
Two types of density matrix $D$ and $\tilde{D}$ are considered,
corresponding to {\it strict} and {\it relaxed} constraints,
respectively. It is notable that that $\tilde{D}$ satisfies
\begin{eqnarray}
N_e = {\rm Tr}\left[ \tilde{D}S \right] % (8)
\end{eqnarray}
like Eq. (5) but that the idempotency requirement is relaxed
\begin{eqnarray}
\tilde{D}S\tilde{D}\le\tilde{D} % (9)
\end{eqnarray}
unlike Eq. (6). In ODA, 
$\tilde{D}_{k+1}$ is defined with an interpolation parameter $\lambda\in [0,1]$ as
\begin{eqnarray}
 \tilde{D}_{k+1}=(1-\lambda)\tilde{D}_k+\lambda D_{k+1}
  =\tilde{D}_k + \lambda(D_{k+1}-\tilde{D}_k)
  =\tilde{D}_k + \lambda\Delta D_{k+1}. % (10)
\end{eqnarray}
So, the RHF Fock matrix $F(D)=h+G(D)$ of Eq. (1) is deformed as
\begin{eqnarray}
 \tilde{F}_{k+1} &=& F(\tilde{D}_{k+1}) \nonumber \\
  &=& h+(1-\lambda)G(\tilde{D}_k)+\lambda G(D_{k+1}) 
  =(1-\lambda)\tilde{F}_k+\lambda F_{k+1}. % (11)
\end{eqnarray}
The parameter $\lambda$ is given by minimizing the RHF energy of Eq. (7).
This is simply solved by a line search for minimizer
\begin{eqnarray}
E^{\rm RHF} (\tilde{D}_{k+1})
  = E^{\rm RHF} (\tilde{D}_k) +s\lambda + c\lambda^2, ~~~~ \lambda\in[0,1] % (12)
\end{eqnarray}
where the parameters $s$ and $c$ are given by
\begin{eqnarray}
 s &=& {\rm Tr} \left[ F(\tilde{D}_k)\Delta D_{k+1} \right], \\ % (13)
 c &=& {\rm Tr} 
 \left[ \left\{F(D_{k+1})-F(\tilde{D}_k)\right\}\Delta D_{k+1} \right]. % (14)
\end{eqnarray}
It is easy to find the optimial value of $\lambda$ by a condition
\begin{eqnarray}
\frac{dE^{\rm RHF}}{d\lambda}=s+2c\lambda=0, % (15)
\end{eqnarray}
where $s$ should be negative until converge
in a sense of steepest descendent slope, as carefully discussed in Ref. [29].
Note that $c$ should be positive conversely.
The optimal damping parameter $\lambda\in[0,1]$ can then be given by
\begin{eqnarray}
 \lambda &=& \left\{
	  \begin{array}{ll}
	   1, & {\rm if} ~~ c\le-s/2, \\ % (16)
	   -s/2c, & {\rm otherwise}.
	  \end{array}
	 \right.
\end{eqnarray}
It is notable that ODA does not work in the case of unity in Eq. (16) (see also Eq. (10)).
As a whole, the ODA-based RHF calculations are sketched as below.
\begin{enumerate}
 \item Initialization:
       Choose an initial guess $D_0$. Assemble
       $F_0=F(D_0)$. Compute $E_0=E(D_0)$. Set $\tilde{D}_0=D_0$,
       $\tilde{F}_0=F_0$ and $k=0$.
 \item Iteration:
       \begin{enumerate}
	\item Diagonalize $\tilde{F}_k$ and assemble the density matrix
	      $D_{k+1}$ via the {\it aufbau} principle [9].
	      \label{RHFODA begin}
	\item If the difference $D_{k+1}-D_k$ is enough small then go to
	      termination; the energy convergence should be checked as well.
	\item Assemble the Fock matrix $F_{k+1}=F(D_{k+1})$ and compute the
	      RHF energy $E_{k+1}=E(D_{k+1})$:
	      \[
	        F_{k+1}=h+G(D_{k+1}),
	      \]
	      \[
	       E_{k+1}={\rm Tr}\left[h D_{k+1}+F_{k+1} D_{k+1} \right].
	      \]
	\item Set $\Delta D_{k+1}=\tilde{D}_k-D_{k+1}$.
	\item Compute
	      \[
	      s = {\rm Tr}\left[\tilde{F}_k \Delta D_{k+1}\right],
	      \]
	      \[
	      c = {\rm Tr}\left[\left\{F_{k+1}-\tilde{F}_k\right\}\Delta D_{k+1}\right].
	      \]
	\item Set $\lambda=1$ if $c\le-s/2$ and $\lambda=-s/2c$ otherwise,
	      and interpolate
	      \[
	      \tilde{D}_{k+1}=(1-\lambda)\tilde{D}_k+\lambda D_{k+1},
	      \]
	      \[
	      \tilde{F}_{k+1}=(1-\lambda)\tilde{F}_k+\lambda F_{k+1},
	      \]
	\item Set $k=k+1$ and go to 2(a).
       \end{enumerate}
 \item Termination:
       Set $C=C_{k+1}$, $D=D_{k+1}$, $F=F_{k+1}$ and $E=E_{k+1}$.
\end{enumerate}
Once the early stage of iteration with a reasonable guess for density matrix
has passed successfully, the SCF procedure with
ODA [23] can generally be switched to that with one of DIIS methods [12,13,15,16],
which may be more efficient in accelerations; nonetheless ODA is 
usable for the final convergence, as well. \par

\bigskip
\noindent{\Large\bf 3. Proposal of ODA/UHF} \par
\medskip
\noindent{\large\bf 3.1. Alternate version} \par
The alternate version of UHF/ODA can be regarded as a straightforward
extension of ODA/RHF [23], except for some points addressed later. 
The superscripts of $\alpha$ or $\beta$ for matrices specify the
spin components, according to Ref. [9].  \par

\smallskip
\noindent{\bf 3.1.1. Basic formulation} \par
First, the numbers of $\alpha$-spin and $\beta$-spin electrons are set as
$N_e^\alpha$ and $N_e^\beta$, respectively, for the density matrices.
The $\alpha$ and $\beta$ Fock matrices are defined, respectively, as [9]
\begin{eqnarray}
F^\alpha = h + G'(D^\alpha) + J(D^\beta), \\ % (17)
F^\beta = h + G'(D^\beta) + J(D^\alpha), \\ % (18)
G'(D) = J(D) - K(D), % (19)
\end{eqnarray}
where the two density matrices are given in the same way as Eq. (3)
\begin{eqnarray}
D^\alpha = C^\alpha C^{\alpha *}, \\ % (20)
D^\beta = C^\beta C^{\beta *}. % (21)
\end{eqnarray}
The AO-MO coefficients are obtained by solving the pair of eigenequations
\begin{eqnarray}
F^\alpha C^\alpha = S C^\alpha \varepsilon^\alpha, \\ % (22)
F^\beta C^\beta = S C^\beta \varepsilon^\beta . % (23)
\end{eqnarray}
The UHF electronic energy is then written as
\begin{eqnarray}
 E^{\rm UHF} (D^\alpha,D^\beta) &=&
  {\rm Tr}\left[h(D^{\alpha}+D^{\beta})+\frac{1}{2}G'(D^{\alpha})D^{\alpha}+
   \frac{1}{2}G'(D^{\beta})D^{\beta}+J(D^{\beta})D^{\alpha}\right] \label{UHF Energy}
   \nonumber \\ % 
   &=& \frac{1}{2}
   {\rm Tr}\left[h(D^{\alpha}+D^{\beta})+F(D^{\alpha})D^{\alpha}+F(D^{\beta})D^{\beta}\right], % (24)
\end{eqnarray}
where an equivalence relation due to a classical nature of Coulomb interaction
\begin{eqnarray}
{\rm Tr}\left[J(D^{\beta})D^{\alpha}\right] = 
{\rm Tr}\left[J(D^{\alpha})D^{\beta}\right]  % (25)
\end{eqnarray}
is utilized. \par

For a while, we focus on updating the $\beta$ density matrix, assuming
that the SCF iteration starts on the $\alpha$ Fock matrix construction.
When the {\it relaxed} $\beta$ density matrix incorporating $\lambda^\beta$
is defined as
\begin{eqnarray}
 \tilde{D}_{k+1}^\beta=(1-\lambda^\beta)\tilde{D}_k^\beta+\lambda^\beta D_{k+1}^\beta
 = \tilde{D}_k^\beta + \lambda^\beta \Delta D_{k+1}^\beta, \label{D=(1-a)D+aD} % (26)
\end{eqnarray}
the resulting $\beta$ Fock matrix with the available {\it strict} $D_{k+1}^\alpha$ is given by
\begin{eqnarray}
 \tilde{F}^\beta_{k+1} &=& F(\tilde{D}_{k+1}^\beta,D_{k+1}^\alpha), \\ % (27)
  &=& h+(1-\lambda^\beta)G'(\tilde{D}_{k}^{\beta})+\lambda^\beta G'(D_{k+1}^{\beta})
    +J(D_{k+1}^{\alpha}), \\ % (28)
  &=& (1-\lambda^\beta)\left[h+G'(\tilde{D}_{k}^{\beta})+J(D_{k}^{\alpha})\right] \nonumber \\
  &&  +\lambda^\beta\left[h+G'(D_{k+1}^{\beta})+J(D_{k+1}^{\alpha})\right] \nonumber \\
  &&  +(1-\lambda^\beta)\left[J(D_{k+1}^{\alpha})-J(D_{k}^{\alpha})\right], \\ % (29)
  &=&(1-\lambda^\beta)\tilde{F}^\beta_{k}+\lambda^\beta F^\beta_{k+1} \nonumber \\
  && +(1-\lambda^\beta)\left[J(D_{k+1}^{\alpha})-J(D_{k}^{\alpha})\right], % (30)
\end{eqnarray}
and the $\beta$ Coulomb part associated with the next $\alpha$ Fock matrix becomes
\begin{eqnarray}
 \tilde{J}_{k+1}^\beta=J(\tilde{D}^\beta_{k+1})
  =(1-\lambda^\beta)\tilde{J}_k^\beta+\lambda^\beta J_{k+1}^\beta. % (31)
\end{eqnarray}
The UHF energy minimizer of interest is then defined as
\begin{eqnarray}
 E^{\rm UHF} (D_{k+1}^\alpha,\tilde{D}_{k+1}^\beta)
  =E^{\rm UHF} (D_{k+1}^\alpha,\tilde{D}_k^\beta) +s^\beta \lambda^\beta + c^\beta (\lambda^\beta)^2 ,
  ~~~~ \lambda^\beta\in[0,1]. \label{AltUHFODA E=E+sx+cx^2} % (32)
\end{eqnarray}
The crucial parameters $s^\beta$ and $c^\beta$ are here given by
\begin{eqnarray}
 s^\beta &=& {\rm Tr}\left[ F(\tilde{D}_k^\beta,D_{k+1}^\alpha)\Delta D_{k+1}^\beta \right], \nonumber \\
   &=& {\rm Tr}\left[\left\{\tilde{F}^\beta_{k+1}+J_{k+1}^\alpha-
   J_k^\alpha\right\}\Delta D_{k+1}^\beta\right], \\ % (33)
 c^\beta &=& \frac{1}{2}{\rm Tr}\left[\left\{F(D_{k+1}^\beta,D_{k+1}^\alpha)
   -F(\tilde{D}_k^\beta,D_{k+1}^\alpha)\right\}\Delta D_{k+1}^\beta\right], \nonumber \\
   &=& \frac{1}{2}{\rm Tr}
   \left[\left\{F^\beta_{k+1}-\tilde{F}_k^\beta
   -J_{k+1}^\alpha+J_{k}^\alpha\right\}\Delta D_{k+1}^\beta\right] % (34)
\end{eqnarray}
with the relation
\begin{eqnarray}
F(D_k^\beta,D_{k+1}^\alpha)=F^\beta_k+J_{k+1}^\alpha-J_k^\alpha . % (35)
\end{eqnarray}
As in the case of ODA/RHF described in the previous SubSection, 
the condition
\begin{eqnarray}
\frac{dE^{\rm UHF}}{d\lambda^\beta}=s^\beta+2c^\beta\lambda^\beta=0, % (36)
\end{eqnarray}
leads to an optimal $\lambda^\beta $ which minimizes the UHF energy function of Eq. (32),
and the result is shown as
\begin{eqnarray}
 \lambda^\beta &=& \left\{
	  \begin{array}{ll}
	   1, & {\rm if} ~ |c^\beta| \le |s^\beta|/2 ~{\rm or}~ s^\beta c^\beta \ge 0 \\ % (37)
	   -s^\beta/2c^\beta, & {\rm otherwise}.
	  \end{array}
	 \right.
\end{eqnarray}
The conditions for $s^\beta$ and $c^\beta$ are slightly modified from those of ODA/RHF [29], since
the simple assumption of a steepest descendent search for the RHF energy is not valid
for the alternate UHF calculations due to the stepwise coupling with the $\alpha$ matrices. \par

\smallskip
\noindent{\bf 3.1.2. Algorithmic flow} \par
The above-mentioned way to derive $\lambda^\beta$ is applicable to $\lambda^\alpha$ as well.
The ODA/UHF procedure for alternate Fock matrix constructions can now be stated as follows.
\begin{enumerate}
 \item Initialization:
       Choose initial guesses $D_0^\alpha$ and 
       $D_0^\beta = D_{-1}^\beta $. Assemble
       $F_0^\alpha=F(D_0^\alpha,D_0^\beta)$,
       $F_0^\beta=F(D_0^\beta,D_0^\alpha)$,
       $J_0^\alpha=J(D_0^\alpha)$
       and $J_0^\beta=J(D_0^\beta)$.
       Compute $E_0=E(D_0^\alpha,D_{-1}^\beta)$.
       Set $\tilde{D}_0^\alpha=D_0^\alpha$,
       $\tilde{F}_0^\alpha=F_0^\alpha$, 
       $\tilde{D}_{-1}^\beta=D_0^\beta$,
       $\tilde{F}_{-1}^\beta=F_0^\beta$,
       $\Delta D_0^\beta=0$ and $k=0$.
 \item Iteration:
       \begin{enumerate}
	\item Diagonalize $\tilde{F}_k^\alpha$ and assemble the density matrix
	      $D_{k+1}^\alpha$ via the {\it aufbau} principle.
	      \label{aUHFODA begin}
	\item Assemble the Fock matrix $F_{k}^\beta$ and the Coulomb integrals $J_{k+1}^\alpha$:
	      \begin{eqnarray*}
	        J_{k+1}^\alpha&=&J(D_{k+1}^\alpha),\\
	        F_{k}^\beta&=&h+G'(D_k^\beta)+J(D_{k+1}^\alpha).
	      \end{eqnarray*}
	\item Set $\Delta D_{k+1}^\alpha=\tilde{D}_k^\alpha-D_{k+1}^\alpha$.
	\item Compute
	      \begin{eqnarray*}
              s^\beta &=& {\rm Tr}\left[\left\{\tilde{F}_{k-1}^\beta-J_k^\alpha+J_{k+1}^\alpha\right\}\Delta D_k^\beta\right],\\
	      c^\beta &=& \frac{1}{2}{\rm Tr}\left[\left\{F_{k}^\beta-\tilde{F}_{k-1}^\beta-J_k^\alpha+J_{k+1}^\alpha\right\}\Delta D_k^\beta\right].
	      \end{eqnarray*}
	\item Set $\lambda^\beta=1$ if $|c^\beta| \le |s^\beta|/2$ or $s^\beta c^\beta \ge 0$ and
              $\lambda^\beta=-s^\beta/2c^\beta$ otherwise,
	      and interpolate
	      \begin{eqnarray*}
	      \tilde{D}_{k}^\beta&=&(1-\lambda^\beta)\tilde{D}_{k-1}^\beta+\lambda^\beta D_{k}^\beta,\\
	      \tilde{F}_{k}^\beta&=&(1-\lambda^\beta)\tilde{F}_{k-1}^\beta+\lambda^\beta
	      F_{k}^\beta + (1-\lambda^\beta)(J_{k+1}^\alpha-J_k^\alpha).
	      \end{eqnarray*}
	\item Diagonalize $\tilde{F}_{k}^\beta$ and assemble the density matrix
	      $D_{k+1}^\beta$ via {\it aufbau} principle.
	\item Assemble the Fock matrix $F_{k+1}^\alpha$ and the Coulomb
	      integrals $J_{k+1}^\beta$:
	      \begin{eqnarray*}
	        J_{k+1}^\beta&=&J(D_{k+1}^\beta),\\
	        F_{k+1}^\alpha&=&h+G'(D_{k+1}^\alpha)+J(D_{k+1}^\beta).
	      \end{eqnarray*}	      
	\item Set $\Delta D_{k+1}^\beta=\tilde{D}_k^\beta-D_{k+1}^\beta$.
	\item Compute the UHF energy
	      \[
	       E_{k+1}=\frac{1}{2}{\rm Tr}\left[hD_{k+1}^\alpha+hD_{k+1}^\beta+F_{k+1}^\alpha
	      D_{k+1}^\alpha+F_{k+1}^\beta D_{k+1}^\beta\right].
	      \]
	\item If the differences $D_{k+1}^\alpha-D_k^\alpha$ and
	      $D_{k+1}^\beta-D_k^\beta$ are enough small then go to termination;
              the energy convergence should be checked as well.
	\item Compute
	      \begin{eqnarray*}
              s^\alpha &=& {\rm Tr}\left[\left\{\tilde{F}_k^\alpha-J_k^\beta+J_{k+1}^\beta\right\}\Delta D_{k+1}^\alpha\right],\\
	      c^\alpha &=& \frac{1}{2}{\rm Tr}\left[\left\{F_{k+1}^\alpha-\tilde{F}_k^\alpha-J_k^\beta+J_{k+1}^\beta\right\}\Delta D_{k+1}^\alpha\right].
	      \end{eqnarray*}
	\item Set $\lambda^\alpha=1$ if $|c^\alpha| \le |s^\alpha|/2$ or $s^\alpha c^\alpha \ge 0$ and 
              $\lambda^\alpha=-s^\alpha/2c^\alpha$ otherwise,
              and interpolate
	      \begin{eqnarray*}
	      \tilde{D}_{k+1}^\alpha&=&(1-\lambda^\alpha)\tilde{D}_k^\alpha+\lambda^\alpha D_{k+1}^\alpha,\\
	      \tilde{F}_{k+1}^\alpha&=&(1-\lambda^\alpha)\tilde{F}_k^\alpha+\lambda^\alpha
	      F_{k+1}^\alpha + (1-\lambda^\alpha)(J_{k+1}^\beta-J_k^\beta).
	      \end{eqnarray*}
	\item Set $k=k+1$ and go to 2(a).
       \end{enumerate}
 \item Termination:
       Set  $C^\alpha=C_{k+1}^\alpha$, $C^\beta=C_{k+1}^\beta$
       $D^\alpha=D_{k+1}^\alpha$, $D^\beta=D_{k+1}^\beta$, $F^\alpha=F_{k+1}^\alpha$,
       $F^\beta=F_{k+1}^\beta$ and $E=E_{k+1}$.
\end{enumerate}
As can be seen above, the computational cost of the 
alternate version of ODA/UHF is roughly twice that of ODA/RHF
when the integral-direct processing [25] is pursued. \par

\medskip
\noindent{\large\bf 3.2. Concurrent version} \par
In the concurrent UHF calculation, both $D^\alpha$ and $D^\beta$ are simultaneously
updated in a certain step of SCF iterations. The concurrent ODA/UHF procedure
is rather complicated in comparison with the alternate ODA/UHF just shown.
This complexity is attributed to the two dimensional nature of minimization algorithm 
instead of one dimensional line search algorithm in the case of alternate version. \par

\smallskip
\noindent{\bf 3.2.1. Basic formulation} \par
A couple of Fock matrices including both {\it relaxed} $\alpha$ and $\beta$
density matrices are defined as (refer also to Eq. (27) for the alternate case)
\begin{eqnarray}
\tilde{\tilde{F}}^\alpha=F(\tilde{D}^\alpha,\tilde{D}^\beta)
=h+G'(\tilde{D}^\alpha)+J(\tilde{D}^\beta) , \\ % (38)
\tilde{\tilde{F}}^\beta=F(\tilde{D}^\beta,\tilde{D}^\alpha)
=h+G'(\tilde{D}^\beta)+J(\tilde{D}^\alpha) ,  % (39)
\end{eqnarray}
and the updates of density matrices in a step of iteration are done as (see Eq. (26))
\begin{eqnarray}
 \tilde{D}_{k+1}^\alpha&=&
  (1-\lambda^\alpha)\tilde{D}_k^\alpha+
  \lambda^\alpha D_{k+1}^\alpha
  = \tilde{D}_k^\alpha + \lambda^\alpha\Delta D_{k+1}^\alpha,
  \ \lambda^\alpha\in[0,1], \\ % (40)
 \tilde{D}_{k+1}^\beta&=&
  (1-\lambda^\beta)\tilde{D}_k^\beta+
  \lambda^\beta D_{k+1}^\beta
   = \tilde{D}_k^\beta + \lambda^\beta\Delta D_{k+1}^\beta,
  \ \lambda^\beta\in[0,1]. % (41)
\end{eqnarray}
The $\alpha$ Fock matrix is then calculated as
\begin{eqnarray}
 \tilde{\tilde{F}}^\alpha_{k+1}
  &=&F(\tilde{D}_{k+1}^\alpha,\tilde{D}_{k+1}^\beta), \\ % (42)
  &=&
   h+(1-\lambda^\alpha)G'(\tilde{D}_{k}^{\alpha})+\lambda^\alpha G'(D_{k+1}^{\alpha})
    +(1-\lambda^\beta )J(\tilde{D}_{k}^{\beta} )+\lambda^\beta  J(D_{k+1}^{\beta}), \\ % (43)
  &=&
   (1-\lambda^\alpha)\left[h+G'(\tilde{D}_{k}^{\alpha})+J(\tilde{D}_{k}^{\beta})\right]
   +\lambda^\alpha \left[h+G'(D_{k+1}^{\alpha})+J(D_{k+1}^{\beta})\right] \nonumber \\
  &&  +(\lambda^\beta-\lambda^\alpha)\left[J(D_{k+1}^{\beta})-J(\tilde{D}_{k}^{\beta} )\right], \\ % (44)
  &=&(1-\lambda^\alpha)\tilde{\tilde{F}}^\alpha_{k}+\lambda^\alpha F^\alpha_{k+1}
   +(\lambda^\beta-\lambda^\alpha)\left[J^\beta_{k+1}-\tilde{J}_k^\beta\right], % (45)
\end{eqnarray}
and the final expression of the $\beta$ Fock matrix becomes
\begin{eqnarray}
 \tilde{\tilde{F}}^\beta_{k+1}
  &=&F(\tilde{D}_{k+1}^\beta,\tilde{D}_{k+1}^\alpha), \\ % (46)
  &=&(1-\lambda^\beta)\tilde{\tilde{F}}^\beta_{k}+\lambda^\beta F^\beta_{k+1}
     +(\lambda^\alpha-\lambda^\beta)\left[J^\alpha_{k+1}-\tilde{J}_k^\alpha\right]. % (47)
\end{eqnarray}
The Coulomb matrices are derived as (see Eq. (31))
\begin{eqnarray}
 \tilde{J}_{k+1}^\alpha&=&J(\tilde{D}_{k+1}^\alpha)=
  (1-\lambda^\alpha)\tilde{J}_k^\alpha+
  \lambda^\alpha J_{k+1}^\alpha, \\ % (48)
 \tilde{J}_{k+1}^\beta&=&J(\tilde{D}_{k+1}^\beta)=
  (1-\lambda^\beta)\tilde{J}_k^\beta+
  \lambda^\beta J_{k+1}^\beta.  % (49)
\end{eqnarray}
The UHF energy minimizer is set with an $\alpha$-$\beta$ coupling term as
\begin{equation}
  E^{\rm UHF} (\tilde{D}_{k+1}^\alpha,\tilde{D}_{k+1}^\beta)
  =E^{\rm UHF} (\tilde{D}_{k}^\alpha,\tilde{D}_k^\beta)
  + s^\alpha\lambda^\alpha
  + s^\beta\lambda^\beta
  + c^\alpha(\lambda^\alpha)^2
  + c^\beta(\lambda^\beta)^2
  + t\lambda^\alpha\lambda^\beta , 
  \label{uhf-oda-obj-func} % (50)
\end{equation}
where the five crucial parameters $s^\alpha$, $s^\beta$, $c^\alpha$, $c^\beta$ and $t$
are obtained as
\begin{eqnarray}
 s^\alpha &=& {\rm Tr}\left[\tilde{\tilde{F}}^\alpha_k\Delta D^\alpha_{k+1}\right], \\ % (51)
 s^\beta &=& {\rm Tr}\left[\tilde{\tilde{F}}^\beta_k\Delta D^\beta_{k+1}\right], \\ % (52)
 c^\alpha &=& \frac{1}{2}
  {\rm Tr}\left[G(\Delta D^\alpha_{k+1})\Delta D^\alpha_{k+1}\right]
  =\frac{1}{2}
  {\rm Tr}\left[\left(F^\alpha_{k+1}-\tilde{\tilde{F}}^\alpha_{k}\right)
	      \Delta D^\alpha_{k+1}\right]
  -\frac{1}{2}t, \\ % (53)
 c^\beta &=& \frac{1}{2}
  {\rm Tr}\left[G(\Delta D^\beta_{k+1})\Delta D^\beta_{k+1}\right]
  =\frac{1}{2}
  {\rm Tr}\left[\left(F^\beta_{k+1}-\tilde{\tilde{F}}^\beta_{k}\right)
	      \Delta D^\beta_{k+1}\right]
  -\frac{1}{2}t, \\ % (54)
 t &=& {\rm Tr}\left[J(\Delta D^\beta_{k+1})\Delta D^\alpha_{k+1}\right], \nonumber \\
   &=& {\rm Tr}\left[\left(\tilde{J}^\beta_k -J^\beta_{k+1} \right)\Delta D^\alpha_{k+1}\right]
    =  {\rm Tr}\left[\left(\tilde{J}^\alpha_k-J^\alpha_{k+1}\right)\Delta D^\beta_{k+1}\right]. % (55)
\end{eqnarray}
In the above derivation, an equivalence relation of Eq. (25) is used. 
The $\alpha$-$\beta$ coupling via $t$ should be effective in accelerating the SCF convergence,
as pointed out for a second-order optimization of UHF in Ref. [36]. Although 
the optimal damping factors $\lambda^\alpha$ and 
$\lambda^\beta$ may be formally determined by these parameters
through the respective partial differentiations, some more consideration is necessary. \par

\smallskip
\noindent{\bf 3.2.2. Two dimensional Newton problem} \par
Eq. (50) can be rewritten as a second-order expansion with respect to $\lambda^\alpha$ and
$\lambda^\beta$
\begin{eqnarray}
 E^{\rm UHF}_{k+1}(\hat{\lambda}) = 
 E^{\rm UHF}_{k}(0) + \hat{g}^* \hat{\lambda}
 + \frac{1}{2} \hat{\lambda}^* \hat{H} \hat{\lambda}, \label{uhf-obj-func-lambda} % (56)
\end{eqnarray}
where the hat is to indicate the two dimensional vectors and matrix
\begin{eqnarray}
\hat{\lambda} = \left(\begin{array}{cc} \lambda^\alpha \\ \lambda^\beta \end{array} \right), \\ % (57)
\hat{g}=\frac{\partial E^{\rm UHF}_{k+1}}{\partial \hat{\lambda}} 
       = \left(\begin{array}{cc} s^\alpha \\ s^\beta \end{array} \right), \\ % (58)
\hat{H}
  = \frac{\partial^2 E^{\rm UHF}_{k+1}}{\partial \hat{\lambda} \partial \hat{\lambda}}
  = \left(\begin{array}{cc} 2c^\alpha & t \\ t & 2c^\beta \end{array} \right). % (59)
\end{eqnarray}
The minimization of Eq. (56) leads to a simple Newton problem
\begin{eqnarray}
\hat{H}\hat{\lambda} = - \hat{g} % (60)
\end{eqnarray}
as long as the Hessian is positive-definite.
If this assumption is valid, the descendent $\lambda$ vector is obtained as 
\begin{eqnarray}
 \left(\begin{array}{cc} \lambda^\alpha \\ \lambda^\beta \end{array} \right)
 = - \hat{H}^{-1} \hat{g} = - \frac{1}{ 4c^\alpha c^\beta - t^2 }
 \left(\begin{array}{cc} 2c^\beta & -t \\ - t & 2c^\alpha \end{array} \right)
 \left(\begin{array}{cc} s^\alpha \\ s^\beta \end{array} \right).  % (61)
\end{eqnarray}
The final expressions for $\lambda^\alpha$ and $\lambda^\beta$ thus become
\begin{eqnarray}
\lambda^\alpha = \frac{2s^\alpha c^\beta - t s^\beta}{t^2 - 4c^\alpha c^\beta} , \\ % (62)
\lambda^\beta =  \frac{2s^\beta c^\alpha - t s^\alpha}{t^2 - 4c^\alpha c^\beta},  % (63)
\end{eqnarray}
under the condition of $\lambda^\alpha,\lambda^\beta\in[0,1]$. Note that the neglection of $t$
yields the essentially same expression as in the case of alternate ODA/UHF for each spin;
this could actually result in a slow convergence by our experiences.
When $\lambda^\alpha,\lambda^\beta\notin[0,1]$, the values should be set as unity to disable ODA. \par

\smallskip
\noindent{\bf 3.2.3. Modified Newton problem} \par
In the two dimensional problem of optimization, the Hessian of Eq. (59) is not restricted to 
be positive definite unfortunately. Namely, the two eigenvalues
\begin{eqnarray}
 \sigma_\pm = (c_\alpha+c_\beta)\pm\sqrt{(c_\alpha-c_\beta)^2+t^2},\label{eigen-value of H} % (64)
\end{eqnarray}
can take three cases (i) positive definite $(0<\sigma_-\le\sigma_+)$, (ii) saddle point
$(\sigma_-\le 0<\sigma_+)$, and (iii) non-positive definite $(\sigma_-\le \sigma_+ \le 0)$.
For cases (ii) and (iii), the technique of shifted Hessian [29,37,38] is usable as
(compare with Eq. (60))
\begin{eqnarray}
(\hat{H}+\mu\hat{1}) \underline{\hat{\lambda}} = - \hat{g}, % (65)
\end{eqnarray}
where $\mu$ for unit matrix $\hat{1}$ is the shift parameter set latter. The modified
solution of $\underline{\hat{\lambda}}$ are then obtained as
\begin{eqnarray}
\underline{\lambda}^{\alpha}
= \frac{(2c^\beta+\mu) s^\alpha  - t s^\beta}{t^2 - (2c^\alpha + \mu) (2c^\beta+\mu)} , \\ % (66)
\underline{\lambda}^{\beta} 
= \frac{(2c^\alpha+\mu) s^\beta - t s^\alpha}{t^2 - (2c^\alpha + \mu) (2c^\beta+\mu)} .  % (67)
\end{eqnarray}
Although the direction of $\underline{\hat{\lambda}}$ should be adjusted
by this modification, two issues still remain. First, $\underline{\hat{\lambda}}$
is not the solution of Eq. (60) manifestly. 
Second, the length of $\underline{\hat{\lambda}}$ may still override the correct 
region of $[0,1]$. These difficulties can be avoided by introducing a scaling relation [37]
\begin{eqnarray}
 \hat{\lambda} = \zeta \underline{\hat{\lambda}} , ~~\hat{\lambda} \in[0,1], % (68)
\end{eqnarray}
and the minimization problem of Eq. (56) is rewritten as
\begin{eqnarray}
 E^{\rm UHF}_{k+1}(\lambda(\zeta)) = 
 E^{\rm UHF}_{k}(0) + \zeta \hat{g}^* \underline{\hat{\lambda}}
 + \frac{1}{2} \zeta^2 \underline{\hat{\lambda}}^* \hat{H} \underline{\hat{\lambda}}, 
 ~~ \zeta\in[0,1] . \label{uhf-obj-func-lambda'}  % (69)
\end{eqnarray}
The $\zeta$ is then obtained as
\begin{eqnarray}
 \zeta =
  \left\{
   \begin{array}{cl}
    1, & {\rm if}~\underline{\hat{\lambda}}^* \hat{H} \underline{\hat{\lambda}} \le
         - \hat{g}^* \underline{\hat{\lambda}} , \\
    - \hat{g}^* \underline{\hat{\lambda}}/
      \underline{\hat{\lambda}}^* \hat{H} \underline{\hat{\lambda}},
	 & {\rm otherwise}. % (70)
   \end{array}
 \right.\label{alpha}
\end{eqnarray}
The scaling factor $\zeta$ can be regarded as a second damping factor consequently.
Anyhow, the relation of $\hat{\lambda} \in[0,1]$ as interpolation factors
should be maintained. Finally, the shift parameter $\mu$ is set as
\begin{eqnarray}
 \mu=\left\{
      \begin{array}{cl}
       0 & {\rm for}~ 0 < \sigma_- \le \sigma_+ , \\
       (\sigma_+ - \sigma_-)/2 & {\rm for}~ \sigma_-\le 0 < \sigma_+ , \\
       -\sigma_- & {\rm for}~ \sigma_- \le \sigma_+ \le 0 , \\ % (71)
      \end{array}
     \right.
\end{eqnarray}
in the actual processing. 

\smallskip
\noindent{\bf 3.2.4. Algorithmic flow} \par
The concurrent ODA/UHF calculations can be described as follows.
\begin{enumerate}
 \item Initialization:
       Choose an initial guess $D_0^\alpha$ and $D_0^\beta$.
       Assemble $F_0^\alpha=F(D_0^\alpha,D_0^\beta)$, 
       $F_0^\beta=F(D_0^\beta,D_0^\alpha)$,
       $J_0^\alpha=J(D_0^\alpha)$ and $J_0^\beta=J(D_0^\beta)$.
       Compute $E_0=E(D_0^\alpha,D_0^\beta)$. Set
       $\tilde{D}_0^\alpha=D_0^\alpha$, $\tilde{D}_0^\beta=D_0^\beta$,
       $\tilde{\tilde{F}}_0^\alpha=F_0^\alpha$, $\tilde{\tilde{F}}_0^\beta=F_0^\beta$,
       $\tilde{J}_0^\alpha=J_0^\alpha$, $\tilde{J}_0^\beta=J_0^\beta$
       and $k=0$.
 \item Iteration:
       \begin{enumerate}
	\item Diagonalize $\tilde{\tilde{F}}_k^\alpha$, 
                          $\tilde{\tilde{F}}_k^\beta$ and assemble
	      $D_{k+1}^\alpha$, $D_{k+1}^\beta$ via the {\it aufbau} principle.
	      \label{cUHFODA begin}
	\item Assemble Fock matrices $F_{k+1}^\alpha$ and $F_{k+1}^\beta$
	      as well as Coulomb integrals $J_{k+1}^\alpha$ and $J_{k+1}^\beta$,
	      \begin{eqnarray*}
	        J_{k+1}^\alpha&=&J(D_{k+1}^\alpha),\\
	        J_{k+1}^\beta&=&J(D_{k+1}^\beta),\\
	        F_{k+1}^\alpha&=&h+G'(D_{k+1}^\alpha)+J(D_{k+1}^\beta),\\
	        F_{k+1}^\beta&=&h+G'(D_{k+1}^\beta)+J(D_{k+1}^\alpha).
	      \end{eqnarray*}
	\item Compute the UHF energy
	      \[
	       E_{k+1}=\frac{1}{2}{\rm Tr}\left[hD_{k+1}^\alpha+hD_{k+1}^\beta+F_{k+1}^\alpha
	       D_{k+1}^\alpha+F_{k+1}^\beta D_{k+1}^\beta\right].
	      \]
	\item If the differences $D_{k+1}^\alpha-D_k^\alpha$ and
	      $D_{k+1}^\beta-D_k^\beta$ are enough small then go to 
	      termination; the energy convergence should be checked as well.
	\item Set
	      $\Delta D_{k+1}^\alpha=\tilde{D}_k^\alpha-D_{k+1}^\alpha$ and
	      $\Delta D_{k+1}^\beta=\tilde{D}_k^\beta-D_{k+1}^\beta$.
	\item Compute
	      \begin{eqnarray*}
	      s^\alpha &=& {\rm Tr}\left[\tilde{\tilde{F}}^\alpha_k \Delta D^\alpha_{k+1}\right],\\
	       s^\beta  &=& {\rm Tr}\left[\tilde{\tilde{F}}^\beta_k  \Delta D^\beta_{k+1}\right],\\
	      c^\alpha &=&
		\frac{1}{2}
		{\rm Tr}\left[\left( F^\alpha_{k+1}-\tilde{\tilde{F}}^\alpha_{k} \right)
			    \Delta D^\alpha_{k+1} \right]
		-\frac{1}{2}t, \\
	       c^\beta &=&
		\frac{1}{2}
		{\rm Tr}\left[\left( F^\beta_{k+1}-\tilde{\tilde{F}}^\beta_{k}\right)
			    \Delta D^\beta_{k+1}\right]
		-\frac{1}{2}t,\\
	       t  &=&
		{\rm Tr}\left[\left(\tilde{J}^\alpha_k-J^\alpha_{k+1}\right)\Delta D^\beta_{k+1}\right].
	      \end{eqnarray*}
      	\item Compute eigenvalues:
	      \begin{eqnarray*}
	       \sigma_\pm = (c^\alpha+c^\beta)\pm\sqrt{(c^\alpha-c^\beta)^2+t^2}.
	      \end{eqnarray*}
	\item Set a shift parameter $\mu$ as follows: 
	      \begin{eqnarray*}
	       \mu=\left\{
		    \begin{array}{cl}
		     0 & {\rm for}~ 0<\sigma_-\le\sigma_+,\\
		     (\sigma_+-\sigma_-)/2 & {\rm for}~ \sigma_-\le0<\sigma_+,\\
		     -\sigma_- & {\rm for}~ \sigma_-\le\sigma_+\le0.
		    \end{array}
		   \right.
	      \end{eqnarray*}
	\item Set a set of tentative damping factors 
              $(\underline{\lambda}^\alpha,\underline{\lambda}^\beta)=(1,1)$ if
	      $\underline{\lambda}^\alpha\notin[0,1]$ or $\underline{\lambda}^\beta\notin[0,1]$ and
	      \begin{eqnarray*}              
              \underline{\lambda}^{\alpha}
              = \frac{(2c^\beta+\mu) s^\alpha  - t s^\beta}{t^2 - (2c^\alpha + \mu) (2c^\beta+\mu)} , \\
              \underline{\lambda}^{\beta}  
              = \frac{(2c^\alpha+\mu) s^\beta - t s^\alpha}{t^2 - (2c^\alpha + \mu) (2c^\beta+\mu)}.
	      \end{eqnarray*}
              otherwise.
	\item Set an optimal scaling factor $\zeta=1$ if
	      $\underline{\hat{\lambda}}^* \hat{H} \underline{\hat{\lambda}} \le -\hat{g}^* \underline{\hat{\lambda}}$ and
	      $\zeta=-\hat{g}^* \underline{\hat{\lambda}}/
               \underline{\hat{\lambda}}^* \hat{H} \underline{\hat{\lambda}}$ otherwise,
	      where
	      \begin{eqnarray*}
              \hat{g}= \left(\begin{array}{cc} s^\alpha \\ s^\beta \end{array} \right) ~{\rm and}~
              \hat{H}= \left(\begin{array}{cc} 2c^\alpha & t \\ t & 2c^\beta \end{array} \right).
	      \end{eqnarray*}
	\item Set the fimal set of dampling factors $\lambda^\alpha=\zeta\underline{\lambda}^\alpha$ and
	      $\lambda^\beta=\zeta\underline{\lambda}^\beta$, and compute interpolations: 
	      \begin{eqnarray*}
	       \tilde{D}_{k+1}^\alpha
		&=&
		(1-\lambda^\alpha)\tilde{D}_k^\alpha+
		\lambda^\alpha D_{k+1}^\alpha,
		\\
	       \tilde{D}_{k+1}^\beta
		&=&
		(1-\lambda^\beta)\tilde{D}_k^\beta+
		\lambda^\beta D_{k+1}^\beta,
		\\
	       \tilde{\tilde{F}}^\alpha_{k+1}
	       &=&(1-\lambda^\alpha)\tilde{\tilde{F}}^\alpha_{k}+\lambda^\alpha F^\alpha_{k+1}
		+(\lambda^\beta-\lambda^\alpha)\left[J^\beta_{k+1}-\tilde{J}_k^\beta\right],
		\\
	       \tilde{\tilde{F}}^\beta_{k+1}
	       &=&(1-\lambda^\beta)\tilde{\tilde{F}}^\beta_{k}+\lambda^\beta F^\beta_{k+1}
		+(\lambda^\alpha-\lambda^\beta)\left[J^\alpha_{k+1}-\tilde{J}_k^\alpha\right],
		\\
	       \tilde{J}_{k+1}^\alpha
		&=&
		(1-\lambda^\alpha)\tilde{J}_k^\alpha+
		\lambda^\alpha J_{k+1}^\alpha,\\
	       \tilde{J}_{k+1}^\beta
		&=&
		(1-\lambda^\beta)\tilde{J}_k^\beta+
		\lambda^\beta J_{k+1}^\beta.
	      \end{eqnarray*}
	\item Set $k=k+1$ and go to 2(a)
       \end{enumerate}
 \item Termination:
       Set $C^\alpha=C_{k+1}^\alpha$, $C^\beta=C_{k+1}^\beta$,
       $D^\alpha=D_{k+1}^\alpha$, $D^\beta=D_{k+1}^\beta$, $F^\alpha=F_{k+1}^\alpha$,
       $F^\beta=F_{k+1}^\beta$ and $E=E_{k+1}$.
\end{enumerate}
As just seen, the concurrent algorithm is more complicated than the alternate one.

\bigskip
\noindent{\Large\bf 4. Test calculations} \par
To test the proposed ODA algorithms of both alternate and concurrent
UHF calculations, we implemented them into a local version
of ABINIT-MPX [39], our original program for the fragment molecular
orbital (FMO) calculations [40], to which UHF energy and its nuclear 
gradient had been implemented [41] as an independent work from Ref. [42].
We here performed the regular UHF (without FMO) calculations for four small 
radicals of the spin doublet (or single open-shell). The 6-31G${}^*$ basis set [43] 
was used for CN, NO${}_2$ and (H${}_2$O)${}_3$+OH. 
A hexa-aqua divalent copper complex, Cu${}^{+2}$+(H${}_2$O)${}_6$,
was calculated with the 6-31G basis set [43], where the D${}_{\rm 2h}$ symmetry 
was imposed for the Jahn-Teller deformation due to 3d${}^9$ occupation. 
The geometries of four molecular systems were optimized by the GAUSSIAN03 
program [44] at the UHF level.
The extended H\"uckel method was used to guess the initial values 
of AO-MO coefficients or density matrices. The convergence
conditions of SCF iterations (cycle limit 1000) were tightly set as 
$\|E_{k}-E_{k-1}\|<10^{-8}$, $\|D_{k}-D_{k-1}\|<10^{-8}$ (for occupied MOs) 
and ${\rm Max}(D_kSF_{k-1}-F_{k-1}SD_k)<10^{-6}$. For testing purpose, 
we enforced the ODA/UHF procedure throughout
(although C2-DIIS [15] was available as well). The reference UHF energies and
spin expectation values computed by GAUSSIAN03 were reproduced by
ABINIT-MPX within reasonable numerical tolerance when converged. 
For comparison, the simple SCF procedure (in the sense of Roothaan)
was adopted as well by the fixed setting $\lambda^\alpha = \lambda^\beta = 1$. \par

Figure 1 plots the convergence behaviors of the CN calculations.
The alternate UHF procedure both with and 
without ODA shows smooth convergence, where no acceleration is 
obtained due to a continuous resetting of the damping parameter as unity 
during the iteration. The concurrent UHF calculation without ODA fails
in convergence, as expected from a demanding nature of this radical.
In contrast, the concurrent ODA/UHF provides a convergence comparable
to the alternate treatment. For NO${}_2$ presented in Figure 2, 
the concurrent calculation without ODA has a slow convergence, 
and the ODA acceleration works well. The behavior of the alternate 
calculations is similar to the case of CN. \par

As seen in Figure 3, the energy lowering with the concurrent ODA
procedure is rapid in early steps for (H${}_2$O)${}_3$+OH.
However, its acceleration drops off near the convergence in
six decimal places unfortunately. This suggests that
the acceleration procedure is to be switched to other methods
such as DIIS [12,13,15,16] once certain criteria of initial 
convergence are passed. Note that the resetting of damping parameter 
took place for the concurrent calculation after the early stage. \par

Figure 4 shows that both ODAs converged
for Cu${}^{+2}$+(H${}_2$O)${}_6$ but the SCF iterations without ODA
lead to the oscillation. Notably, the concurrent version is 
much better, especially in the early stage. As denoted in the above paragraph, 
ODA should be switched to DIIS for the accelerated final convergence
in production runs. We tried another initial guess with the diagonalization of $h$ 
(core Hamiltonian). As a result of this attempt, the concurrent ODA calculation converged 
as in the case of H\"uckel guess, while the alternate one oscillated (data not shown). 
A notable merit of ODA/RHF could be a robustness against poor initial 
guesses [23], and this might be valid also for the concurrent ODA/UHF 
procedure in which the $\alpha$-$\beta$ coupling is carefully taken 
into account. Finally, it is a favorable fact that the concurrent version 
works better than does the alternate one for the four examples 
employed here, since the the former can be faster in 
processing of AO-integrals [25]. \par

\bigskip
\noindent{\Large\bf 5. Summary} \par
In this paper, we proposed two ODAs for UHF calculations of open-shell
molecular systems, as extensions of the original ODA/RHF developed by Canc\`es and Bris [23].
The equations associated with the Fock and density matrices were systematically 
derived for both alternate and concurrent SCF procedures.
In the latter procedure, an additional two-dimensional Newton method was 
employed to determine the optimal set of damping factors. 
Test calculations were performed for four doublet
radical systems. It was shown that the concurrent ODA has better
overall performance in convergence than does the alternate one.
This fact should be favorable since the number of integral 
evaluations could be halved in integral-direct SCF computations [25,26].
Works to fully implement the proposed recipes are underway for
the improved performance of FMO-UHF calculations in the ABINIT-MPX program [39,41]. \par

\bigskip
\noindent{\Large\bf Acknowledgements} \par
This work was supported by the SFR-aid by Rikkyo University and the RISS project
at the Institute of Industrial Science (IIS) of the University of Tokyo. We owe
the first implementation of UHF calculations (without the treatment of
$\alpha$-$\beta$ coupling term) in ABINIT-MPX to Mr. Yuji Kato.
Finally, the authors thank Dr. Yuto Komeiji for comments on the manuscript. \par

\newpage

\noindent{\Large\bf References} \par
% \begin{thebibliography}{44} % reference list for UHF-ODA paper - 2013/1/19
\begin{enumerate}

% --- RHF, UHF and ROHF ---

\item % \bibitem{Roothaan1951} % 1 / SCF procedure of RHF
Roothaan CCJ (1951) Rev Mod Phys 23:69

\item % \bibitem{Pople1954} % 2 / SCF procedure of UHF
Pople JA, Nesbet RK (1954) J Chem Phys 22:571

\item % \bibitem{Roothaan1960} % 3 / SCF procedure of ROHF
Roothaan CCJ (1960) Rev Mod Phys 32:179

\item % \bibitem{Plakhutin2006} % 4 / SCF procedure of ROHF - canonical Fock
Plakhutin BN, Gorelik EV, Breslavskaya NN (2006) J Chem Phys 125:204110

\item % \bibitem{Tuchimochi2010} % 5 / UHF-ROHF relation revival 1
Tsuchimochi T, Henderson TM, Scuseria GE, Savin A (2010) J Chem Phys 133:134108

\item % \bibitem{Tuchimochi2010} % 6 / UHF-ROHF relation revival 2
Tsuchimochi T, Scuseria GE (2010) J Chem Phys 133:141102

% --- Acceleration method for RHF-SCF ---

\item % \bibitem{Koutecky1975} % 7 / Convergence difficulty - old 2 (open)
Sleeman DH (1968) Theor Chim Acta 11:135

\item % \bibitem{Koutecky1971} % 8 / Convergence difficulty - old 1 (closed)
Kouteck\'y J, Bona\( \check{\rm c} \)i\'c V (1971) J Chem Phys 55:2408

\item % \bibitem{Szabo1982} % 9 / General textbook
Szabo A, Ostlund NS (1982) Modern Quantum Chemistry. MacMillan, New York

\item % \bibitem{Sanuders1973} % 10 / Level shift for SCF
Saunders VR, Hiller IH (1973) Intern J Quant Chem 7:699

\item % \bibitem{Zerner1979} % 11 / Dynamic damping
Zerner MC, Hehenberger M (1979) Chem Phys Lett 62:550

\item % \bibitem{Pulay1980} % 12 / DIIS-SCF 1
Pulay P (1980) Chem Phys Lett 73:393

\item % \bibitem{Pulay1982} % 13 / DIIS-SCF 2
Pulay P (1982) J Comp Chem 3:556

\item % \bibitem{Sellers1991} % 14 / Improved DIIS - ADEM-DIOS
Sellers H (1991) Chem Phys Lett 180:461

\item % \bibitem{Sellers1993} % 15 / Improved DIIS - C2
Sellers H (1993) Intern J Quant Chem 45:31

\item % \bibitem{Kudin2002} % 16 / EDIIS <Sub-main paper>
Kudin KK, Scuseria GE, Canc\`es E (2002) J Chem Phys 116:8255

\item % \bibitem{Host2008} % 17 / Augmented RH and DIIS
H$\o$st S, Olsen J, Jans\'ik B, Th$\o$gersen L, J$\o$rgensen P, 
Helgaker T (2008) J Chem Phys 129:124106

\item % \bibitem{Hu2010} % 18 / ADIIS
Hu X, Yang W (2010) J Chem Phys 132:054109

\item % \bibitem{Wang2011} % 19 / LISTi - Listd
Wang YA, Yam CY, Chen YK, Chen G (2011) J Chem Phys 134:241103

\item % \bibitem{Chen2011} % 20 / LISTb
Chen YK, Wang YA (2011) J Chem Theor Comp 7:3045

\item % \bibitem{Chen2011} % 21 / DIIS comparison
Garza AJ, Scuseria GE (2012) 137:054110

\item % \bibitem{Canses2000} % 22 / Preparation for ODA
Canc\`es E, Bris CL (2000) Math Model Num Anal 34:749

\item % \bibitem{Canses2000} % 23 / ODA for RHF <Main paper>
Canc\`es E, Bris CL (2000) Intern J Quant Chem 79:82

\item % \bibitem{Bacskay1981} % 24 / QC-SCF - MO
Bacskay GB (1981) Chem Phys 61:385

\item % \bibitem{Almlof1982} % 25 / Direct SCF - DISCO
Alml\( \ddot {\rm o} \)f J, Faegri K, Korsell K (1982) 
J Comp Chem 3:385

\item % \bibitem{Feyereisen1993} % 26 / Parallel DISCO
Feyereisen M, Kendall RA (1993) Theor Chim Acta 84:289

\item % \bibitem{Shepard1993} % 27 / SO-SCF - Fock - MO
Shepard R (1993) Theor Chim Acta 84:343

\item % \bibitem{Rendell1994} % 28 / SO-SCF - Fock - MO
Rendell AP (1994) Chem Phys Lett 229:204

\item % \bibitem{Wong1995} % 29 / SO-SCF - Fock - MO
Wong AT, Harrison RJ (1995) J Comp Chem 16:1291

\item % \bibitem{Chaban1997} % 30 / SO-SCF (UHF,CAS) - Fock - MO
Chaban G, Schmidt MW, Gordon MS (1997) Theor Chem Acc 97:88

\item % \bibitem{Mochizuki2005} % 31 / CERF - MO
Mochizuki Y (2005) Chem Phys Lett 410:165

\item % \bibitem{Mochizuki2005} % 32 / AO-based Newton TR-SCF
Sa$\l$ek P, H$\o$st S, Th$\o$gersen L, J$\o$rgensen P, Manninen P,
Olsen J, Jans\'ik B, Reine S, Paw$\l$owski F, Tellgren E, Helgaker T,
Coriani S (2007) J Chem Phys 126:114110

% --- Acceleration method for UHF-SCF ---

\item % \bibitem{Claxton1971} % 33 / UHF energy minimization
Claxton TA, Smith NA (1971) Theor Chim Acta 22:399

\item % \bibitem{Claxton1971} % 34 / UHF energy minimization
Seeger R, Pople JA (1976) J Chem Phys 65:265

\item % \bibitem{Bacskay1982} % 35 / QC-SCF - Open-Shell (ROHF) MO
Bacskay GB (1982) Chem Phys 65:383

\item % \bibitem{Neese2000} % 36 / SO-SCF - Open-shell (UHF) MO
Neese F (2000) Chem Phys Lett 325:93

% --- Level-shifted Newton ---

\item % \bibitem{Shepard1982} % 37 / MCSCF Newton step
Shepard R, Shavitt I, Simons J (1982) J Chem Phys 76:543

\item % \bibitem{Jensen1984} % 38 / MCSCF Newton step
Jensen HJA, J$\o$rgensen P (1984) J Chem Phys 80:1204

% --- Remaining ---

\item % \bibitem{Nakano2012} % 39 / ABINIT-MPX
Mochizuki Y, Yamashita K, Fukuzawa K, Takematsu K, Watanabe H, 
Taguchi N, Okiyama Y, Tsuboi M, Nakano T, Tanaka S (2010) 
Chem Phys Lett 493:346

\item % \bibitem{Kitaura1999} % 40 / FMO 1st paper
Kitaura K, Ikeo E, Asada T, Nakano T, Uebayasi M (1999) Chem Phys Lett 313:701

\item % \bibitem{Kato2013} % 41 / FMO3-UHF and gradient - MD
Kato Y, Komeiji Y, Fujiwara T, Nakano T, Mori H, Yamamoto J,
Mochizuki Y (to be published)

\item % \bibitem{Nakata2012} % 42 / FMO2-UHF and gradient
Nakata H, Fedorov DG, Nagata T, Yokogawa S, Ogata K, Kitaura K,
Nakamura S (2012) J Chem Phys 137:044110

\item % \bibitem{GAUSSIAN_Book1996} % 43 / Gaussian basis sets
Foresman JB, Frisch A (1996)
Exploring Chemistry with Electronic Structure Methods (2nd Ed). 
Gaussian Inc., Pittsburgh

\item % \bibitem{GAUSSIAN03} % 44 / Geom. Opt. program and reference energy
GAUSSIAN03 (Rev. D.02) (2003) Gaussian Inc. Pittsburgh, http://www.gaussian.com

% \end{thebibliography}
\end{enumerate}

\newpage
\noindent{\Large\bf Figure captions} \par
\noindent {\bf Figure 1.} Convergence behavior of CN molecule
($ {}^2 \Sigma^+ $ state).
E${}_{\rm c}$ is the finally converged UHF energy, and
E${}_{\rm k}$ means the snapshot energies during the SCF iteration. The result of
concurrent UHF calculation with ODA is labeled as "Conc/ODA" (red solid line),
while the case without ODA is plotted as "Conc/SCF" 
(purple broken line). The behaviors of alternate UHF calculations 
with and without ODA are shown with labels of "Alter/ODA" (blue broken line) 
and "Alter/SCF" (green dotted line), respectively. \par \bigskip

\noindent {\bf Figure 2.} Convergence behavior of NO${}_2$ molecule
($ {}^2 A_1 $ state). The captions are the same as those of Figure 1. \par \bigskip

\noindent {\bf Figure 3.} Convergence behavior of (H${}_2$O)${}_3$+OH
cluster ($ {}^2 A $ state). The captions are the same as those of Figure 1. \par \bigskip

\noindent {\bf Figure 4.} Convergence behavior of Cu${}^{+2}$+(H${}_2$O)${}_6$
complex ($ {}^2 A_{\rm g} $ state).
The captions are the same as those of Figure 1. \par \bigskip

\newpage
\begin{figure}
 \begin{center}
 \includegraphics*{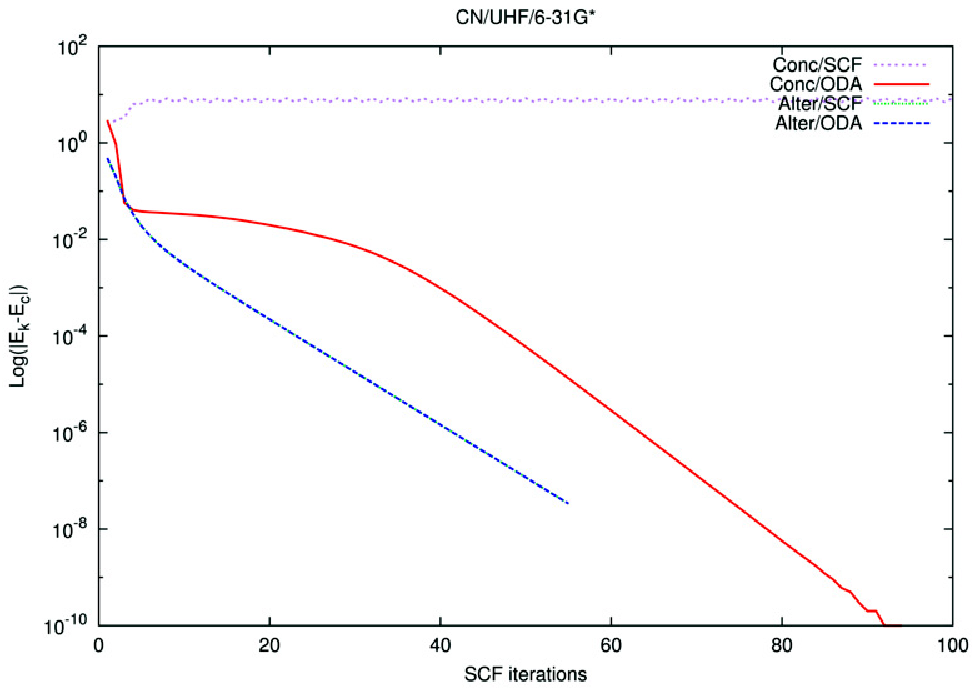}
 \end{center}
\end{figure}

\newpage
\begin{figure}
 \begin{center}
 \includegraphics*{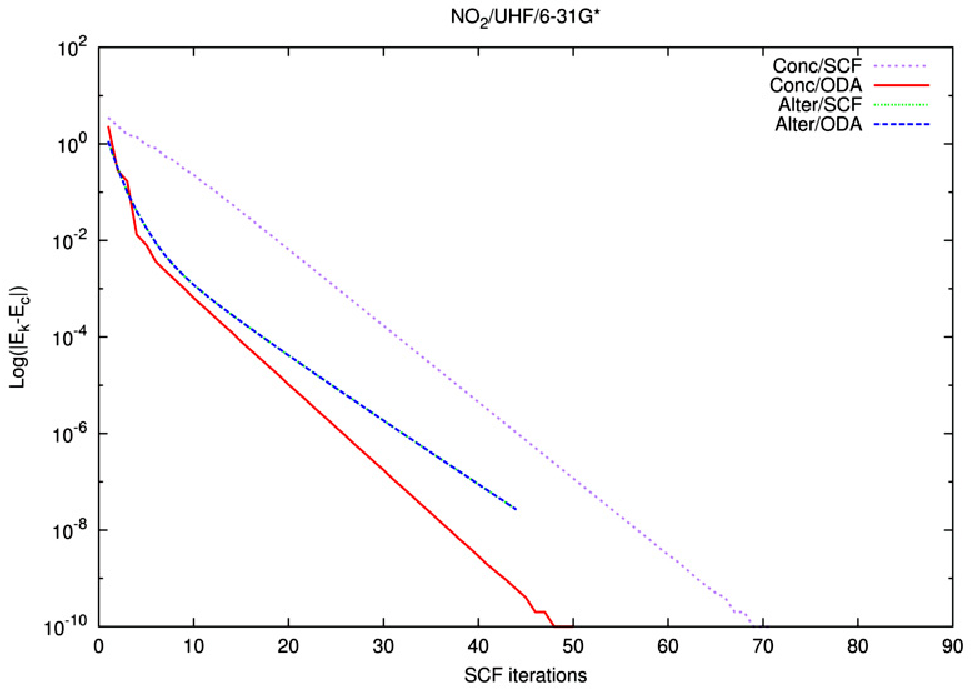}
 \end{center}
\end{figure}

\newpage
\begin{figure}
 \begin{center}
 \includegraphics*{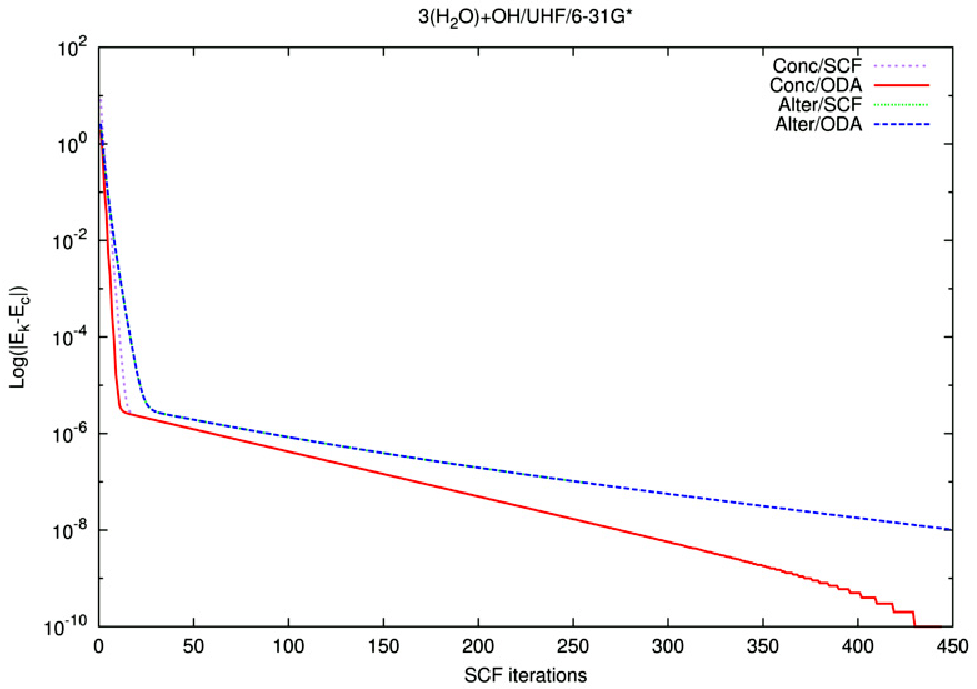}
 \end{center}
\end{figure}

\newpage
\begin{figure}
 \begin{center}
 \includegraphics*{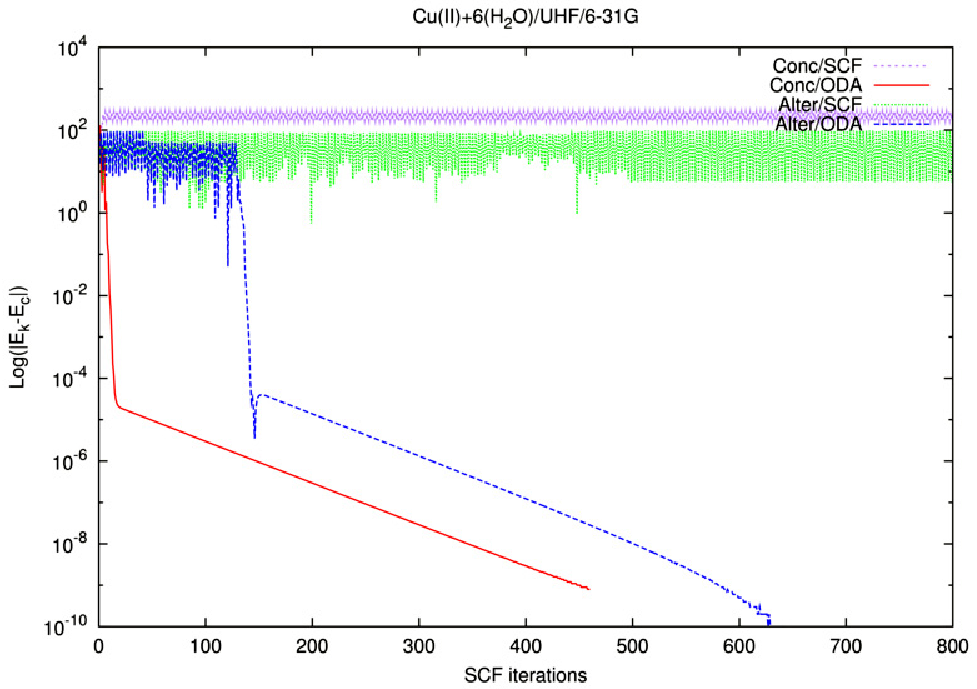}
 \end{center}
\end{figure}

\end{document}